\documentclass{ws-ijqi_}

\newcommand{\bra}[1]{\langle #1|}
\newcommand{\ket}[1]{|#1 \rangle}

\newcommand{\oP}[2]{\ket{#1}\bra{#2}}

\begin{document}

\title{A Novel Attack Strategy on Entanglement Swapping QKD Protocols}

\author{Stefan Schauer}
\address{
Department of Quantum Technology, Austrian Research Centers GmbH - ARC\\
Lakeside B.01, Klagenfurt 9020, Austria\\
Stefan.Schauer@arcs.ac.at}

\author{Martin Suda}
\address{Department of Quantum Technology, Austrian Research Centers GmbH - ARC\\
Donau-City-Str. 1, Vienna 1220, Austria\\
Martin.Suda@arcs.ac.at}

\maketitle

\begin{abstract}
Li et al. presented a protocol [Int. Journal of Quantum Information, Vol. 4, No. 6 (2006) 899-906] 
for quantum key distribution based on entanglement swapping. In this protocol they use random and 
certain bits to construct a classical key and they claim that this key is secure. In our article we 
show that the protocol by Li et al. is insecure presenting a new type of attack strategy which gives 
an adversary full information about the key without being detected. This strategy is based on 
entanglement swapping, too, and manages to preserve the correlation between the measurement results 
of the legitimate parties. Further we present a modified version of the protocol and show that it is 
secure against this new attack strategy.
\end{abstract}


\section{Introduction} \label{sec_Intro}

Entanglement swapping \cite{ZZHE93} is a phenomenon where 2 or more qubits which didn't interact in 
the past are brought into an entangled state. For example, two parties Alice and Bob share the 
Bell states $\ket{\Phi^+}_{12}$ and $\ket{\Phi^+}_{34}$ such that Alice is in possession of qubits 1 
and 3 and Bob of qubits 2 and 4. If Alice performs a measurement in the Bell basis on qubits 1 and 3 
she will end up with one of the four Bell states. Simultaneously Bob's qubits 2 and 4 are also 
brought into a Bell state. This measurement can be described by the equation
\begin{eqnarray}
\ket{\Phi^+}_{12} \otimes \ket{\Phi^+}_{34}
= &\dfrac{1}{2} \Bigl(&
\ket{\Phi^+}_{13} \ket{\Phi^+}_{24} + \ket{\Phi^-}_{13} \ket{\Phi^-}_{24} \nonumber \\ 
&+&\ket{\Psi^+}_{13} \ket{\Psi^+}_{24} + \ket{\Psi^-}_{13} \ket{\Psi^-}_{24} \Bigr)
\label{eq_EntSwap}
\end{eqnarray}
This equation shows that Alice's result is completely random, i.e. she will obtain any of the four 
Bell states with equal probability. But the state of the qubits 2 and 4 is completely determined by 
the result of Alice's measurement. This correlation is used in a number of key distribution 
protocols \cite{WZT05,Cab01,LSZW03,GGWZ05,Cab00,Son04,Cab00_2} to share a secret key between Alice 
and Bob. Since the practical realization of entanglement swapping is rather complex to achieve with 
today's technology none of these QKD protocols has been implemented yet.
\par
Also Li et al. presented a QKD protocol \cite{LWWSZ06} based on entanglement swapping which not only 
produces random key bits but also certain key bits. This is achieved by introducing a Pauli 
operation in eq. (\ref{eq_EntSwap}) as it will be described in detail in section 
\ref{sec_OrigProt}. As it is shown in \refcite{LWWSZ06} this protocol is secure against an 
intercept/resend attack as well as a collective attack. But we will present an attack strategy in 
section \ref{sec_AttStrat} which is based on a 6-qubit state. This state is more complex than the 
system of the sender and the receiver, but it provides an adversary with full information about the 
key shared between the two parties.
\par
The main idea of this attack is that the adversary Eve entangles herself with Alice and Bob using a 
state which preserves the correlations of Alice's Bell state measurement. Further this state 
contains additional information which allows Eve to eavesdrop Bob's secret result. An attack where
Eve is entangled with one party was presented by Zhang, Li and Guo \cite{ZLG00_1} on the protocol
of Cabello \cite{Cab00_2} but it wouldn't work for the protocol in \refcite{LWWSZ06} since Eve 
wouldn't stay undetected.
\par
We will also make a proposal how to secure the protocol against this new attack strategy. This can 
be achieved using a Hadamard operation and we will describe the new version in detail in section 
\ref{sec_ModProt}. In section \ref{sec_SecAnal} we will discuss the security of our new version.

\section{The Original Protocol} \label{sec_OrigProt}

Alice creates $2n$ EPR pairs, each in a Bell state, e.g. $\ket{\Phi^+}$. In the following we will 
focus on the simple case $n=1$, where Alice and Bob share the states $\ket{\Phi^+}_{12}$ and 
$\ket{\Phi^+}_{34}$, which are publicly known. In \refcite{LWWSZ06} Li et al. start the protocol with 
preshared entangled states which is a very strong assumption because the qubits have to be shared 
somehow between the two parties. But in the security analysis in \refcite{LWWSZ06} an adversary Eve is
granted the possibility to interact with the Bell states. Thus we can conclude that there have to be 
some qubits in transit between Alice and Bob. There are two possible ways to share the Bell states 
between the legitimate parties: either one party, e.g. Alice, prepares both Bell states and sends 
two qubits to the other party or both parties prepare one Bell state and they exchange two qubits.
\par
\begin{figure}[pb]
\centerline{\psfig{file=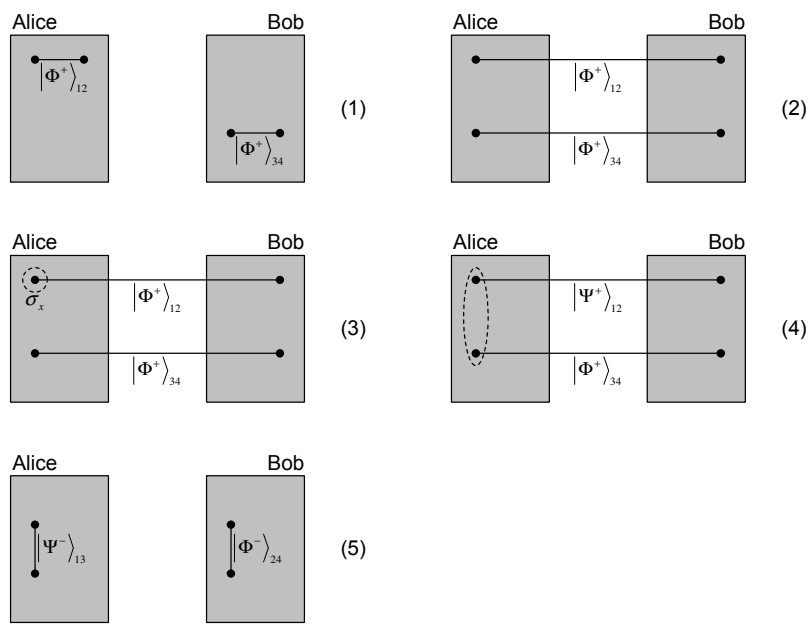,width=0.85\textwidth}}
\vspace*{8pt}
\caption{An illustration of the original protocol by Li at al. $^9$. Here $\sigma_x$ is 
chosen for Alice's secret operation and $\ket{\Psi^-}_{13}$ as Alice's result of the Bell state 
measurement.}
\label{pic_OrigProt}  
\end{figure}

 
The first scenario is insecure against a simple intercept-resend attack: If Alice prepares the Bell 
states $\ket{\Phi^+}_{12}$ and $\ket{\Phi^+}_{34}$ and sends qubits 2 and 4 to Bob Eve can intercept 
these qubits and perform a Bell state measurement on them. This will bring qubits 1 and 3 also into 
a Bell state which is known to Eve. Then she forwards qubits 2 and 4 to Bob. Thus Eve has full 
information about Alice's and Bob's secret measurement result and is able to eavesdrop the key 
perfectly. Therefore we will assume that Alice prepares $\ket{\Phi^+}_{12}$ and Bob prepares
$\ket{\Phi^+}_{34}$ and that they exchange the qubits 2 and 3 (cf. picture (2) in figure 
\ref{pic_OrigProt}).
\par
Alice randomly chooses one of the four Pauli operations, $I$, $\sigma_x$, $\sigma_y$ or $\sigma_z$ 
and applies it to qubit 1 (cf. picture (3) in figure \ref{pic_OrigProt}). Alice's operation will 
generally be described as $\sigma_A^{(i)}$, where the superscript $(i)$ denotes that it is applied 
on qubit $i$. As a next step Alice performs a Bell state measurement on qubits 1 and 3 in her 
possession (cf. picture (4) in figure \ref{pic_OrigProt}).
\par
With this measurement the effect of entanglement swapping arises and entangles Bob's qubits 2
and 4. Following the equation
\begin{eqnarray}
\sigma_A^{(1)} \ket{\Phi^+}_{12} \otimes \ket{\Phi^+}_{34} =
&\dfrac{1}{2} \Bigl(&
\ket{\Phi^+}_{13} \sigma_A^{(2)} \ket{\Phi^+}_{24} + \ket{\Phi^-}_{13} \sigma_A^{(2)} \ket{\Phi^-}_{24} \nonumber \\
&+&\ket{\Psi^+}_{13} \sigma_A^{(2)} \ket{\Psi^+}_{24} + \ket{\Psi^-}_{13} \sigma_A^{(2)} \ket{\Psi^-}_{24} \Bigr)
\label{eq_CESP_ESPO_PD_ESwOp}
\end{eqnarray}
Alice can determine in which state Bob's qubits are after her Bell state measurement if she applied 
the operation $\sigma_A$ on qubit 1. Additionally, she can compute also what the state of qubits 2 
and 4 would be, if she didn't apply any operation on qubit 1 using eq. (\ref{eq_EntSwap}) from 
above. Now Alice announces to Bob that she made a Bell state measurement but keeps her result secret.
\par
To infere Alice's result Bob performs a Bell state measurement on qubit 2 and 4 in his possession. 
Since he does not know yet which operation Alice applied and in which state qubits 1 and 2 have been 
before the measurement he does not know the exact state of qubits 1 and 3. With the help of eq. 
(\ref{eq_EntSwap}) he can at least deduce in which state qubits 1 and 3 should be if Alice 
didn't apply anything to qubit 1. This is called Alice's imaginary result.
\par
To get the correct state of qubits 1 and 3 Bob asks Alice about her result but keeps his own result 
secret. Using the information about Alice's and his own result Bob can deduce which Pauli operation 
Alice applied. Alice and Bob now share information about Alice's randomly chosen operation and the 
result Bob obtained, which is also random. They use these two pieces of information to extract a 
classical raw key. Therefore they agree beforehand on a mapping of Bell states onto classical 2-bit 
strings, e.g.
\begin{equation}
\ket{\Phi^+} \longrightarrow 00 \quad \ket{\Phi^-} \longrightarrow 01 \quad
\ket{\Psi^+} \longrightarrow 10 \quad \ket{\Psi^-} \longrightarrow 11
\label{eq_CESP_ESPO_PD_Map}
\end{equation}
as well as of Pauli operations onto classical 2-bit strings, i.e.
\begin{equation}
I \longrightarrow 00 \quad \sigma_x \longrightarrow 01 \quad
\sigma_y \longrightarrow 10 \quad \sigma_z \longrightarrow 11
\end{equation}
\par
If we take the general case with $2n$ EPR pairs Alice and Bob repeat these steps for all the 
pairs. In the end they publicly compare a certain number of bits of the generated raw key to detect
an eavesdropper.


\section{A Perfect Attack Strategy} \label{sec_AttStrat}

\begin{figure}[pb]
\centerline{\psfig{file=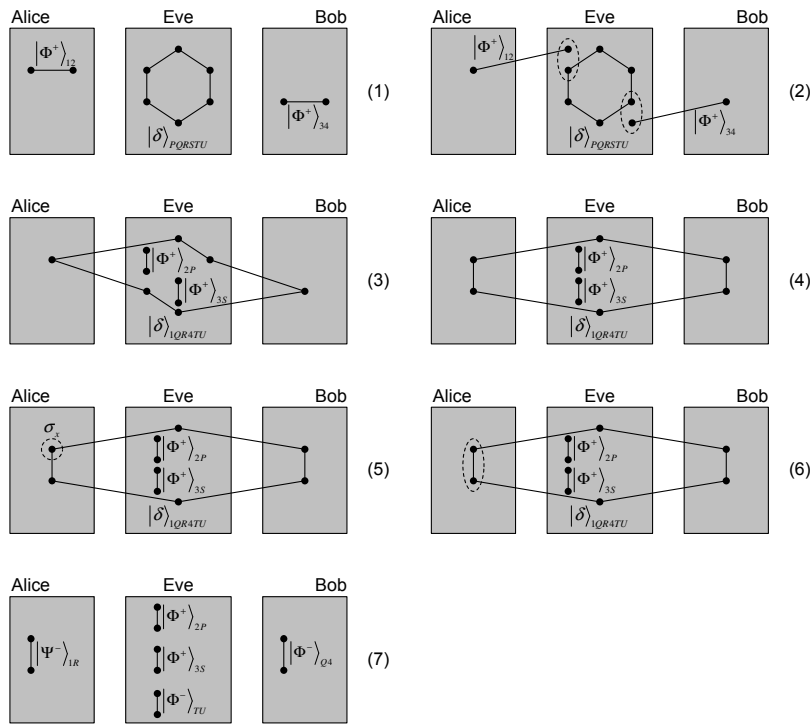,width=0.85\textwidth}}
\caption{An illustration of the new attack strategy described in section \ref{sec_AttStrat}. Again 
$\sigma_x$ is chosen as Alice's secret operation and $\ket{\Psi^-}_{1R}$ as Alice's result of the 
Bell state measurement.}
\label{pic_Attack} 
\end{figure}

In this section we discuss an attack strategy where Eve is able to perfectly eavesdrop the secret 
key of Alice and Bob without being detected. Therefore she uses the six qubit state (cf. picture (1) 
in figure \ref{pic_Attack})
\begin{eqnarray}
\ket{\delta}_{PQRSTU} = &\dfrac{1}{2\sqrt{2}} \Bigl(&
\ket{000000}_{PQRSTU} + \ket{001101}_{PQRSTU} \nonumber \\ 
&+&\ket{010111}_{PQRSTU} + \ket{011010}_{PQRSTU} \nonumber \\
&+&\ket{100110}_{PQRSTU} + \ket{101011}_{PQRSTU} \nonumber \\
&+&\ket{110001}_{PQRSTU} + \ket{111100}_{PQRSTU}
\Bigr)
\label{eq_EveInitState}
\end{eqnarray}
This state is rather complex and not easy to generate (cf. appendix) but it has 
the special property that it can be written as
\begin{eqnarray}
\ket{\delta}_{PQRSTU} = &\dfrac{1}{2} \Bigl(&
\ket{\Phi^+}_{PR} \otimes \ket{\Phi^+}_{QS} \otimes \ket{\Phi^+}_{TU} \nonumber \\ 
&+&\ket{\Phi^-}_{PR} \otimes \ket{\Phi^-}_{QS} \otimes \ket{\Phi^-}_{TU} \nonumber \\
&+&\ket{\Psi^+}_{PR} \otimes \ket{\Psi^+}_{QS} \otimes \ket{\Psi^+}_{TU} \nonumber \\ 
&+&\ket{\Psi^-}_{PR} \otimes \ket{\Psi^-}_{QS} \otimes \ket{\Psi^-}_{TU} 
\Bigr)
\label{eq_EveStateProp}
\end{eqnarray}
That means, a measurement on qubits $P$ and $R$ as well as $Q$ and $S$ yields in the same result as 
it would be expected from entanglement swapping in eq. (\ref{eq_EntSwap}). Further the remaining two 
qubits $T$ and $U$ are in the same state as the qubits $Q$ and $S$. Eve can use this fact to obtain 
enough information to deduce the secret measurement results of Alice and Bob if she somehow manages 
that the two parties share the state $\ket{\delta}$ instead of two Bell states. That can be achieved 
easily using entanglement swapping on qubits 2 and 3 in transit together with her six qubits.
\par
When Alice sends out qubit 2 and Bob qubit 3 Eve preforms Bell state measurements on both 
of them with respective qubits from $\ket{\delta}_{PQRSTU}$ (cf. picture (2) in figure 
\ref{pic_Attack}). A problem is that Eve, at first, doesn't know which initial state Alice prepared. 
But, as we will show in the next paragraphs, she can overcome this problem fairly easy. Eve's first 
Bell state measurement on qubits 2 and $P$ is described in eq. (\ref{eq_CESP_ESPO_AS_ES_ES1}). (As 
already pointed out above, we will use $\sigma_A$ to describe Alice's secret Pauli operation.)
\begin{eqnarray}
\lefteqn{\sigma_A^{(1)} \ket{\Phi^+}_{12} \otimes \ket{\delta}_{PQRSTU} =} \hspace{30pt} \nonumber \\
& &\sigma_A^{(1)} \frac{1}{\sqrt{2}} \Bigl(\ket{00}_{12} + \ket{11}_{12}\Bigr) \otimes
\frac{1}{2\sqrt{2}} \Bigl( \ket{000000}_{PQRSTU} + \ket{001101}_{PQRSTU} \nonumber \\ 
& &\qquad\quad{}+ \ket{010111}_{PQRSTU} + \ket{011010}_{PQRSTU} + \ket{100110}_{PQRSTU} \nonumber \\
& &\qquad\quad{}+ \ket{101011}_{PQRSTU} + \ket{110001}_{PQRSTU} + \ket{111100}_{PQRSTU}\Bigr) \nonumber \\
=&\dfrac{1}{2} \Biggl( &\ket{\Phi^+}_{2P} \otimes \sigma_A^{(1)} \frac{1}{2\sqrt{2}} \Bigl(
\ket{000000}_{1QRSTU} + \ket{001101}_{1QRSTU} \nonumber \\
& &\qquad\quad{}+ \ket{010111}_{1QRSTU} + \ket{011010}_{1QRSTU} + \ket{100110}_{1QRSTU} \nonumber \\
& &\qquad\quad{}+ \ket{101011}_{1QRSTU} + \ket{110001}_{1QRSTU} + \ket{111100}_{1QRSTU} \Bigr) \nonumber \\
&+&\ket{\Phi^-}_{2P} \otimes \sigma_A^{(1)} \frac{1}{2\sqrt{2}} \Bigl(
\ket{000000}_{1QRSTU} + \ket{001101}_{1QRSTU} \nonumber \\
& &\qquad\quad{}+ \ket{010111}_{1QRSTU} + \ket{011010}_{1QRSTU} - \ket{100110}_{1QRSTU} \nonumber \\
& &\qquad\quad{}- \ket{101011}_{1QRSTU} - \ket{110001}_{1QRSTU} - \ket{111100}_{1QRSTU} \Bigr) \nonumber \\
&+&\ket{\Psi^+}_{2P} \otimes \sigma_A^{(1)} \frac{1}{2\sqrt{2}} \Bigl(
\ket{000110}_{1QRSTU} + \ket{001011}_{1QRSTU} \nonumber \\
& &\qquad\quad{}+ \ket{010001}_{1QRSTU} + \ket{011100}_{1QRSTU} + \ket{100000}_{1QRSTU} \nonumber \\
& &\qquad\quad{}+ \ket{101101}_{1QRSTU} + \ket{110111}_{1QRSTU} + \ket{111010}_{1QRSTU} \Bigr) \nonumber \\
&+&\ket{\Psi^-}_{2P} \otimes \sigma_A^{(1)} \frac{1}{2\sqrt{2}} \Bigl(
\ket{000110}_{1QRSTU} + \ket{001011}_{1QRSTU} \nonumber \\
& &\qquad\quad{}+ \ket{010001}_{1QRSTU} + \ket{011100}_{1QRSTU} - \ket{100000}_{1QRSTU} \nonumber \\
& &\qquad\quad{}- \ket{101101}_{1QRSTU} - \ket{110111}_{1QRSTU} - \ket{111010}_{1QRSTU} \Bigr)
\Biggr)
\label{eq_CESP_ESPO_AS_ES_ES1}
\end{eqnarray}
At this point Eve can not clearly say which state the qubits 1, $Q, R, S, T$ and $U$ are in. All 
four results of her Bell state measurement are equally likely and she has absolutely no information 
about the operation $\sigma_A$. Nevertheless, Eve is able to transform the 6-qubit state in a way 
that it is always in the state $\sigma_A^{(1)} \ket{\delta}_{1QRSTU}$. Therefore she performs a 
$\sigma_x$ operation on qubits $S$ and $T$ if she gets $\ket{\Psi^+}_{2P}$ or does nothing if her 
result is $\ket{\Phi^+}_{2P}$. To correct the negative signs in case Eve obtains a 
$\ket{\Phi^-}_{2P}$ she performs a $\sigma_z$ on qubits $S$ and $U$. In case of $\ket{\Psi^-}_{2P}$ 
she first has to apply the $\sigma_x$ operator on qubits $S$ and $T$ and then the $\sigma_z$ 
operator on qubits $S$ and $U$.
\par
The second Bell state measurement is performed on qubits 3 and $S$. Again, the resulting 6-qubit 
state is not entirely known but it can be one out of four possible states. We won't describe this
Bell state measurements explicitly but we want to stress that Eve is able to bring the 6 qubits in 
the state $\sigma_A^{(1)} \ket{\delta}_{1QR4TU}$. Therefore she uses a similar method as described 
above applying the Pauli operations $\sigma_x$ and $\sigma_z$ on qubits $R$, $T$ and $U$.
\par
Now Eve sends qubit $R$ to Alice and qubit $Q$ to Bob and keeps qubits $T$ and $U$ by herself (cf. 
picture (4) in figure \ref{pic_Attack}). Alice and Bob, who are not aware of Eve's intervention 
follow the protocol and perform Bell state measurements on their respective particles (cf. pictures 
(5) and (6) in figure \ref{pic_Attack}). If we look at the 6-qubit state 
$\sigma_A^{(1)} \ket{\delta}_{1QR4TU}$, Alice's Bell state measurement on qubits 1 and $R$ changes 
the state in a way that it leaves the qubit pairs $Q$, 4 and $T$, $U$ in an entangled state. This 
can be seen from the alternative description of the state $\sigma_A^{(1)} \ket{\delta}_{1QR4TU}$ in 
eq. (\ref{eq_CESP_ESPO_AS_ES_ESAlice}) and picture (7) in figure \ref{pic_Attack}).
\begin{eqnarray}
\lefteqn{\sigma_A^{(1)} \ket{\delta}_{1QR4TU} =} \hspace{30pt} \nonumber \\ 
& &\sigma_A^{(1)} \frac{1}{2\sqrt{2}} \Bigl( \ket{000000}_{1QR4TU} + \ket{001101}_{1QR4TU} \nonumber \\
& &\qquad\quad\;{}+ \ket{010111}_{1QR4TU} + \ket{011010}_{1QR4TU} + \ket{100110}_{1QR4TU} \nonumber \\ 
& &\qquad\quad\;{}+ \ket{101011}_{1QR4TU} + \ket{110001}_{1QR4TU} + \ket{111100}_{1QR4TU}\Bigr) \nonumber \\
=& \dfrac{1}{2} \Biggl(&\sigma_A^{(1)} \ket{\Phi^+}_{1R} \otimes \frac{1}{2} \Bigl(
\ket{0000}_{Q4TU} + \ket{0011}_{Q4TU} \nonumber \\ 
& &\qquad\qquad\qquad{}+ \ket{1100}_{Q4TU} + \ket{1111}_{Q4TU} \Bigr) \nonumber \\
&+&\sigma_A^{(1)} \ket{\Phi^-}_{1R} \otimes \frac{1}{2} \Bigl(
\ket{0000}_{Q4TU} - \ket{0011}_{Q4TU} \nonumber \\
& &\qquad\qquad\qquad{}- \ket{1100}_{Q4TU} + \ket{1111}_{Q4TU} \Bigr) \nonumber \\
&+&\sigma_A^{(1)} \ket{\Psi^+}_{1R} \otimes \frac{1}{2} \Bigl(
\ket{0101}_{Q4TU} + \ket{0110}_{Q4TU} \nonumber \\
& &\qquad\qquad\qquad{}+ \ket{1001}_{Q4TU} + \ket{1010}_{Q4TU} \Bigr) \nonumber \\
&+&\sigma_A^{(1)} \ket{\Psi^-}_{1R} \otimes \frac{1}{2} \Bigl(
\ket{0101}_{Q4TU} + \ket{0110}_{Q4TU} \nonumber \\
& &\qquad\qquad\qquad{}- \ket{1001}_{Q4TU} - \ket{1010}_{Q4TU} \Bigr)
\Biggr)
\label{eq_CESP_ESPO_AS_ES_ESAlice}
\end{eqnarray}
Alice gets each possible result with equal probability of $\frac{1}{4}$ as she would expect it.
Further, Bob's Bell state measurement will end up with certainty in a specific Bell state, which is
fully dependent on Alice's result. Due to the special choice of the state $\ket{\delta}$ the 
correlation between his and Alice's result is preserved as they would expect it if no eavesdropper 
was present. Thus the presence of Eve won't be detected when Alice and Bob compare some bits of the 
raw key. Additionally, it leaves the qubits $T$ and $U$, which are in Eve's possession, in the very 
same state Bob obtains as result (cf. also picture (7) in figure \ref{pic_Attack}). This is easy to 
see if we look at the alternative description in the Bell basis of the remaining four qubits $Q$, 4, 
$T$ and $U$ after Alice's measurement
\begin{eqnarray}
\frac{1}{2} \Bigl(
 \sigma_A^{(1)} \ket{\Phi^+}_{1R} \otimes \ket{\Phi^+}_{Q4} \otimes \ket{\Phi^+}_{TU}
&+&\sigma_A^{(1)} \ket{\Phi^-}_{1R} \otimes \ket{\Phi^-}_{Q4} \otimes \ket{\Phi^-}_{TU} \nonumber \\
+\sigma_A^{(1)} \ket{\Psi^+}_{1R} \otimes \ket{\Psi^+}_{Q4} \otimes \ket{\Psi^+}_{TU}
&+&\sigma_A^{(1)} \ket{\Psi^-}_{1R} \otimes \ket{\Psi^-}_{Q4} \otimes \ket{\Psi^-}_{TU}
\Bigr)
\label{eq_CESP_ESPO_AS_ES_ESBob}
\end{eqnarray}
\par
At this time, Eve has full information about Bob's secret measurement. That means, Eve has as much 
information as Bob and thus can obtain the classical secret raw key in the same way Bob does: From 
her result of the Bell state measurement Eve can compute Alice's imaginary result. Moreover, when 
Alice publicly announces her result, Eve is able to infere which Pauli operation Alice has chosen. 
Thus Eve knows both parts of the shared secret information. Therefore she easily can infere the 
classical bit string which is used as a raw key by Alice and Bob and so she is able to obtain the 
secret key, too. 
\par
Instead of performing the Bell state measurements on the qubit pairs 2, $P$ and 3, $S$ as soon as 
she receives them, Eve has the opportunity to delay her measurement and immediately send qubit $R$ 
to Alice (cf. picture (2) in figure \ref{pic_ModProtAttack2}). With this strategy Eve is able to 
overcome operations Alice applies on qubit 1 but she introduces a much higher error rate as we will 
describe in section \ref{sec_SecAnal}. 
Another, but rather strong, assumption is that Eve is in control of Alice's and Bob's EPR source. 
That means she is able to distribute the state $\ket{\delta}_{PQRSTU}$ between her and the two 
parties without further Bell state measurements. In this case Alice's measurement on qubits $P$ and 
$R$ leaves Bob's qubits in a correlated state and Eve with the full information about the results, 
as described above. Since Alice and Bob can prepare their EPR source themself this assumption is not 
very practical.


\section{The Modified Protocol} \label{sec_ModProt}

\begin{figure}[pb]
\centerline{\psfig{file=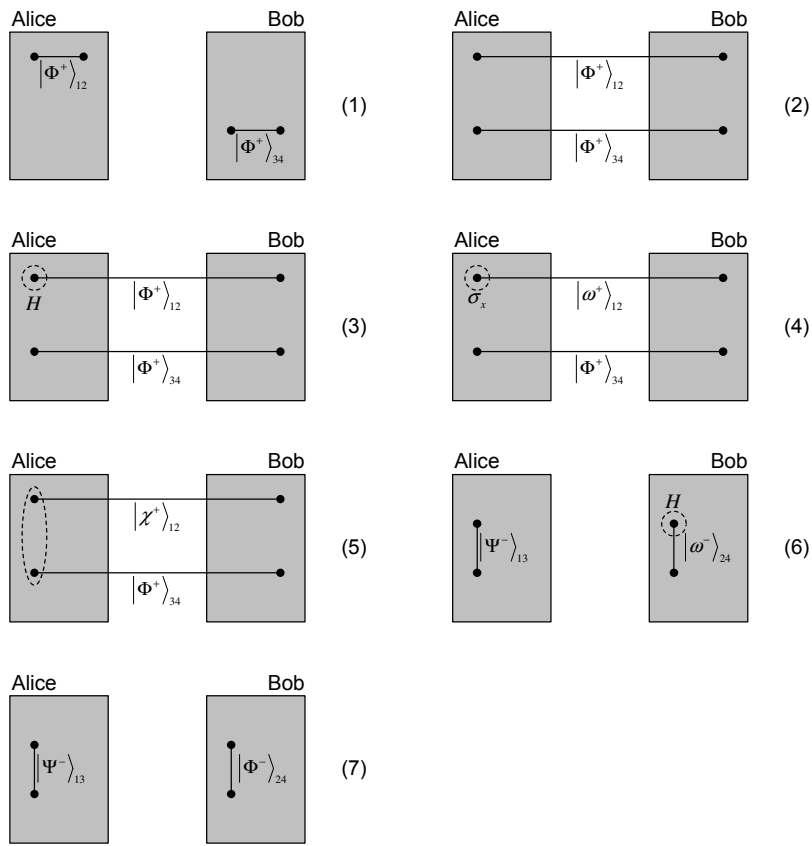,width=0.85\textwidth}}
\caption{An illustration of the modified protocol presented in section \ref{sec_ModProt}. Alice 
performs the Hadamard operation before she applies her secret operation $\sigma_x$. Again, 
$\ket{\Psi^-}_{13}$ is assumed to be Alice's result of the Bell state measurement.}
\label{pic_ModProt}  
\end{figure}

The protocol can be improved by performing a Hadamard operation before Alice applies her secret
Pauli operation on qubit 1 (cf. picture (3) in figure \ref{pic_ModProt}). The Hadamard operation 
alters the initial state in a way that Eve can't eavesdrop the secret key without introducing a 
certain error rate. Thus Alice and Bob can detect her easily as it will be shown in section 
\ref{sec_SecAnal}.
\par
Due to the use of the additional Hadamard operation it is not important any more wether Alice 
prepares both Bell states or each party prepares a Bell state by its own. To stay consistent with 
the above descriptions of the protocol and the attack we will discuss the scenario where Alice 
prepares the state $\ket{\Phi^+}_{12}$ and Bob prepares $\ket{\Phi^+}_{34}$ (for the simple case 
$n=1$). Alice sends out qubit 2 to Bob and he sends qubit 3 to Alice. When she receives Bob's qubit 
she randomly applies either the identity operator $I$ or the Hadamard operator
\begin{equation}
H = \frac{1}{\sqrt{2}} \left( \begin{array}{cc}1 &1 \\ 1 &-1\end{array} \right)
\end{equation}
on qubit 1. This alters the initial state to
\begin{equation}
H^{(1)} \ket{\Phi^+}_{12} \otimes \ket{\Phi^+}_{34} = 
\ket{\omega^+}_{12} \otimes \ket{\Phi^+}_{34}
\label{eq_AltInitState}
\end{equation}
where $\ket{\omega^+}_{12}$ is one of the four possible superpositions of Bell states, i.e.
\begin{eqnarray}
H^{(1)} \ket{\Phi^\pm}_{12} = \ket{\omega^\pm}_{12} &=& 
\frac{1}{\sqrt{2}} \Bigl( \ket{\Phi^\mp}_{12} \pm \ket{\Psi^\pm}_{12} \Bigr) \nonumber \\
H^{(1)} \ket{\Psi^\pm}_{12} = \ket{\chi^\pm}_{12} &=& 
\frac{1}{\sqrt{2}} \Bigl( \ket{\Psi^\mp}_{12} \pm \ket{\Phi^\pm}_{12} \Bigr)
\label{eq_SupBell}
\end{eqnarray}
Then Alice applies her secret Pauli operation $\sigma_A$ on qubit 1 and performs a Bell state 
measurement on qubits 1 and 3 (cf. picture (5) in figure \ref{pic_ModProt}). This Bell state 
measurement can be described as
\begin{eqnarray}
\ket{\omega^+}_{12} \otimes \ket{\Phi^+}_{34} =
&\dfrac{1}{2} \Bigl(& \ket{\Phi^+}_{13} \otimes \ket{\omega^+}_{24}
+\ket{\Phi^-}_{13} \otimes \ket{\omega^-}_{24} \nonumber \\
&+&\ket{\Psi^+}_{13} \otimes \ket{\chi^+}_{24}
+\ket{\Psi^-}_{13} \otimes \ket{\chi^-}_{24} 
\Bigr) 
\end{eqnarray}
Further she publicly tells Bob wether she has applied the $H$ operator or not. If she
did so, Bob applies the $H$ operator on qubit 2 and otherwise he does nothing (cf. picture (6) in 
figure \ref{pic_ModProt}). As we can see from eq. (\ref{eq_SupBell}) a repeated application of the 
Hadamard operator eliminates the superposition and the correlation from the original Bell state 
measurement is preserved.
In the end Bob performs a Bell measurement on qubits 2 and 4 and computes Alice's actual and 
imaginary result as it is described in the original protocol. 

\section{Security Analysis} \label{sec_SecAnal}

\begin{figure}[pb]
\centerline{\psfig{file=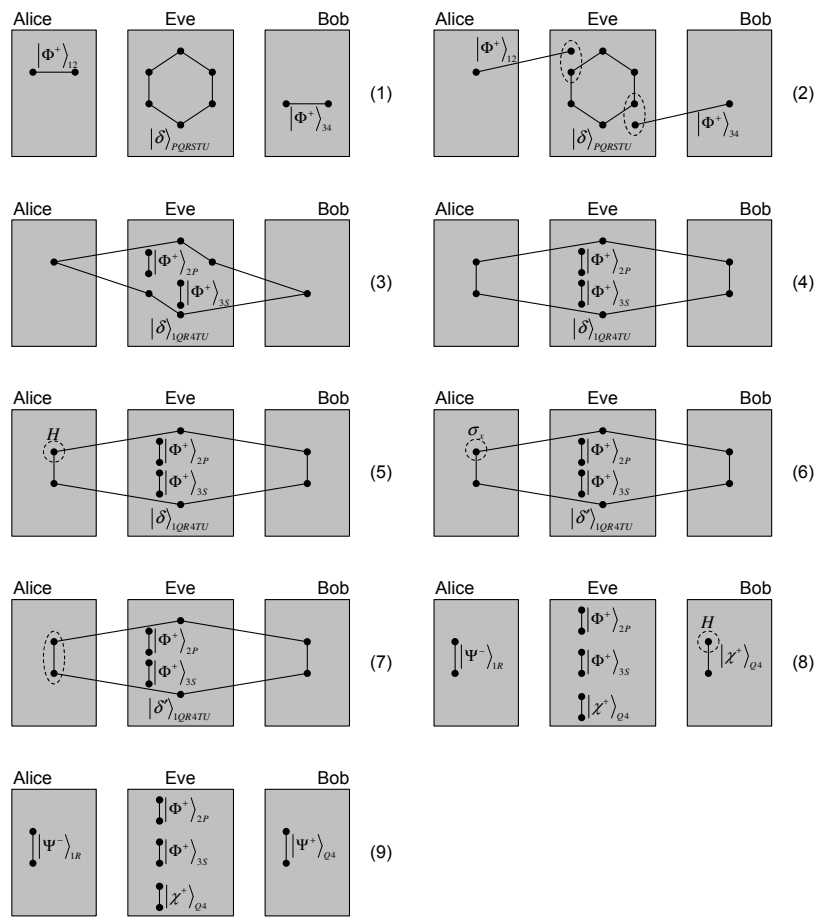,width=0.85\textwidth}}
\caption{An illustration of the attack on the modified protocol as described in section 
\ref{sec_SecAnal}. Eve follows the strategy presented in section \ref{sec_AttStrat} and Alice has 
chosen to apply the Hadamard operation. Again, $\sigma_x$ is Alice's secret operation and 
$\ket{\Psi^-}_{1R}$ is Alice's result of the Bell state measurement.}
\label{pic_ModProtAttack}  
\end{figure}

In the following we will provide a security analysis of the modified protocol. Since the application 
of the Hadamard operation is just a minor change of the original protocol, the modified version is 
also secure against an intercept/resend attack as well as a collective attack, as described in 
\refcite{LWWSZ06}. Therefore we will first inspect Eve's probability to stay undetected if she 
follows the same attack strategy just described in section \ref{sec_AttStrat}. If Alice does not 
apply the Hadamard operator Eve's attack will be successfull as we already pointed out above. Thus 
we will discuss the case where Alice performs the Hadamard operation on qubit 1 (cf. figure 
\ref{pic_ModProtAttack}). Due to the fact that the Hadamard operation alters the initial state as we 
have seen in eq. (\ref{eq_AltInitState}) Eve's first Bell state measurement on qubits 2 and $P$ will 
also change as described in eq. 
(\ref{eq_ModProt_EveBSM}).
\begin{eqnarray}
\lefteqn{\sigma_A^{(1)} H^{(1)} \ket{\Phi^+}_{12} \otimes \ket{\delta}_{PQRSTU} =} \hspace{30pt} \nonumber \\ 
&\dfrac{1}{2} \Biggl(&
\ket{\Phi^+}_{2P} \otimes \sigma_A^{(1)} H^{(1)} \frac{1}{2\sqrt{2}} \Bigl(
\ket{000000}_{1QRSTU} + \ket{001101}_{1QRSTU} \nonumber \\
& &\quad{}+ \ket{010111}_{1QRSTU} + \ket{011010}_{1QRSTU} + \ket{100110}_{1QRSTU} \nonumber \\
& &\quad{}+ \ket{101011}_{1QRSTU} + \ket{110001}_{1QRSTU} + \ket{111100}_{1QRSTU} \Bigr) \nonumber \\
&+&\ket{\Phi^-}_{2P} \otimes \sigma_A^{(1)} H^{(1)} \frac{1}{2\sqrt{2}} \Bigl(
\ket{000000}_{1QRSTU} + \ket{001101}_{1QRSTU} \nonumber \\
& &\quad{}+ \ket{010111}_{1QRSTU} + \ket{011010}_{1QRSTU} - \ket{100110}_{1QRSTU} \nonumber \\
& &\quad{}- \ket{101011}_{1QRSTU} - \ket{110001}_{1QRSTU} - \ket{111100}_{1QRSTU} \Bigr) \nonumber \\
&+&\ket{\Psi^+}_{2P} \otimes \sigma_A^{(1)} H^{(1)} \frac{1}{2\sqrt{2}} \Bigl(
\ket{000110}_{1QRSTU} + \ket{001011}_{1QRSTU} \nonumber \\
& &\quad{}+ \ket{010001}_{1QRSTU} + \ket{011100}_{1QRSTU} + \ket{100000}_{1QRSTU} \nonumber \\
& &\quad{}+ \ket{101101}_{1QRSTU} + \ket{110111}_{1QRSTU} + \ket{111010}_{1QRSTU} \Bigr) \nonumber \\
&+&\ket{\Psi^-}_{2P} \otimes \sigma_A^{(1)} H^{(1)} \frac{1}{2\sqrt{2}} \Bigl(
\ket{000110}_{1QRSTU} + \ket{001011}_{1QRSTU} \nonumber \\
& &\quad{}+ \ket{010001}_{1QRSTU} + \ket{011100}_{1QRSTU} - \ket{100000}_{1QRSTU} \nonumber \\
& &\quad{}- \ket{101101}_{1QRSTU} - \ket{110111}_{1QRSTU} - \ket{111010}_{1QRSTU} \Bigr)
\Biggr)
\label{eq_ModProt_EveBSM}
\end{eqnarray}
As it has been shown in section \ref{sec_AttStrat} Eve is able to change the resulting states of her 
measurements such that she will always end up with a slight variation of her initial state. A 
similar argument holds for Eve's second measurement on qubits 3 and $S$. Thus the state after these
two measurements is (cf. picture (6) in figure \ref{pic_ModProtAttack})
\begin{equation}
\ket{\delta^\prime}_{1QR4TU} = \sigma_A^{(1)} H^{(1)} \ket{\delta}_{1QR4TU}
\label{eq_EveAltState}
\end{equation}
Then Eve sends out qubit $R$ to Alice and qubit $Q$ to Bob. When Alice performs her Bell state 
measurement the state $\ket{\delta^\prime}_{1QR4TU}$ can be described as in eq. 
(\ref{eq_ModProt_AliceBSM}) (cf. also picture (7) in figure \ref{pic_ModProtAttack}).
\begin{eqnarray}
\lefteqn{\sigma_A^{(1)} H^{(1)} \ket{\delta}_{1QR4TU} =} \hspace{30pt} \nonumber \\ 
& &\sigma_A^{(1)} H^{(1)} \frac{1}{2\sqrt{2}} \Bigl( 
  \ket{000000}_{1QR4TU} + \ket{001101}_{1QR4TU} \nonumber \\ 
& &\qquad\quad{}+ \ket{010111}_{1QR4TU} + \ket{011010}_{1QR4TU} + \ket{100110}_{1QR4TU} \nonumber \\
& &\qquad\quad{}+ \ket{101011}_{1QR4TU} + \ket{110001}_{1QR4TU} + \ket{111100}_{1QR4TU}
\Bigr) \nonumber \\
=& \dfrac{1}{2} \Biggl(&
\sigma_A^{(1)} \ket{\Phi^+}_{1R} \otimes \frac{1}{2\sqrt{2}} \Bigl(
\ket{0000}_{Q4TU} - \ket{0011}_{Q4TU} \nonumber \\ 
& &\qquad\quad{}- \ket{1100}_{Q4TU} + \ket{1111}_{Q4TU} + \ket{0101}_{Q4TU} \nonumber \\
& &\qquad\quad{}+ \ket{0110}_{Q4TU} + \ket{1001}_{Q4TU} + \ket{1010}_{Q4TU}
\Bigr) \nonumber \\
&+&\sigma_A^{(1)} \ket{\Phi^-}_{1R} \otimes \frac{1}{2\sqrt{2}} \Bigl(
\ket{0000}_{Q4TU} + \ket{0011}_{Q4TU} \nonumber \\ 
& &\qquad\quad{}+ \ket{1100}_{Q4TU} + \ket{1111}_{Q4TU} - \ket{0101}_{Q4TU} \nonumber \\
& &\qquad\quad{}+ \ket{0110}_{Q4TU} + \ket{1001}_{Q4TU} - \ket{1010}_{Q4TU}
\Bigr) \nonumber \\
&+&\sigma_A^{(1)} \ket{\Psi^+}_{1R} \otimes \frac{1}{2\sqrt{2}} \Bigl(
\ket{0101}_{Q4TU} - \ket{0110}_{Q4TU} \nonumber \\ 
& &\qquad\quad{}- \ket{1001}_{Q4TU} + \ket{1010}_{Q4TU} + \ket{0000}_{Q4TU} \nonumber \\ 
& &\qquad\quad{}+ \ket{0011}_{Q4TU} + \ket{1100}_{Q4TU} + \ket{1111}_{Q4TU}
\Bigr) \nonumber \\
&+&\sigma_A^{(1)} \ket{\Psi^-}_{1R} \otimes \frac{1}{2\sqrt{2}} \Bigl(
\ket{0101}_{Q4TU} + \ket{0110}_{Q4TU} \nonumber \\ 
& &\qquad\quad{}+ \ket{1001}_{Q4TU} + \ket{1010}_{Q4TU} - \ket{0000}_{Q4TU} \nonumber \\
& &\qquad\quad{}+ \ket{0011}_{Q4TU} + \ket{1100}_{Q4TU} - \ket{1111}_{Q4TU}
\Bigr)
\Biggr)
\label{eq_ModProt_AliceBSM}
\end{eqnarray}

For reasons of simplicity we will assume that Alice's result is $\sigma_A^{(1)} \ket{\Phi^-}_{1R}$
as it is illustrated in picture (8) in figure \ref{pic_ModProtAttack} (the computations for any 
other result are analog). Bob also applies the Hadamard operator on qubit $Q$ in his possession and 
thus alters the state of qubits $Q,4,T$ and $U$ to
\begin{eqnarray}
&\dfrac{1}{4} \Bigl(&
   \ket{0000}_{Q4TU} + \ket{1000}_{Q4TU} + \ket{0011}_{Q4TU} + \ket{1011}_{Q4TU} \nonumber \\
&+&\ket{0100}_{Q4TU} - \ket{1100}_{Q4TU} + \ket{0111}_{Q4TU} - \ket{1111}_{Q4TU} \nonumber \\
&-&\ket{0101}_{Q4TU} - \ket{1101}_{Q4TU} + \ket{0110}_{Q4TU} + \ket{1110}_{Q4TU} \nonumber \\ 
&+&\ket{0001}_{Q4TU} - \ket{1001}_{Q4TU} - \ket{0010}_{Q4TU} + \ket{1010}_{Q4TU}
\Bigr)
\end{eqnarray}
which can be alternatively written as
\begin{equation}
\frac{1}{\sqrt{2}} \biggl(
\ket{\Phi^-}_{Q4} \otimes \ket{\omega^-}_{TU} + \ket{\Psi^+}_{Q4} \otimes \ket{\chi^+}_{TU} 
\biggr)
\end{equation}
From this it is easy to see that Bob will get the expected result, $\ket{\Phi^-}_{Q4}$ only with 
50\% probability. In the other half of the cases he will obtain $\ket{\Psi^+}_{Q4}$. Alice and Bob 
will detect this error when they check the correlation between their results.
\par
That means in $\frac{1}{2}$ of the cases, when Alice doesn't apply the Hadamard operation, Eve will 
stay undetected with certainty. In the other half of the cases when Alice performs the Hadamard 
transformation on qubit 1 Eve will be detected with probability $\frac{1}{2}$. These probabilities 
are valid for the simple case ($n=1$) where Alice and Bob share only 2 Bell states. In general, 
Alice and Bob share $2n$ entangled qubit pairs and then they will detect Eve with the probability
\begin{equation}
p = 1 - \biggl( \frac{1}{2} \cdot 1 + \frac{1}{2} \cdot \frac{1}{2} \biggr)^n 
  = 1 - \biggl(\frac{3}{4}\biggr)^n
\label{eq_ModProt_Error}
\end{equation}
which can be brought close to 1 for large $n$.
\par
A possibility for Eve to overcome the effect of the Hadamard operation is to apply a $H$ operation
herself on the qubits $P$ and $Q$ before the protocol starts. This changes the initial state 
$\ket{\delta}_{PQRSTU}$ to $H^{(P)}H^{(Q)}\ket{\delta}_{PQRSTU}$ and in the course of the 
protocol Alice and Bob's application of the Hadamard operator will invert Eve's operations. In 
detail, eq. \ref{eq_EveAltState} changes to
\begin{equation}
\ket{\delta^\prime}_{1QR4TU} = \sigma_A^{(1)} H^{(1)} H^{(1)} H^{(Q)} \ket{\delta}_{1QR4TU}
= \sigma_A^{(1)} H^{(Q)} \ket{\delta}_{1QR4TU}
\end{equation}
and further, after Bob's application of the Hadamard operation
\begin{equation}
H^{(Q)} \ket{\delta^\prime}_{1QR4TU} = \sigma_A^{(1)} H^{(Q)} H^{(Q)} \ket{\delta}_{1QR4TU}
= \sigma_A^{(1)} \ket{\delta}_{1QR4TU}
\end{equation}
which is equal to the state from eq. \ref{eq_CESP_ESPO_AS_ES_ESAlice}. Thus, the attack reduces to 
the version described in section \ref{sec_AttStrat} and Eve will obtain full information about the 
key whenever Alice decides to perform the $H$ operation. In all the other cases Eve will introduce 
an error equal to eq. \ref{eq_ModProt_Error} due to her Hadamard operations, which is easy to see 
following the arguments in the paragraphs above. At this point we want to stress that Alice and Bob 
will become suspicious if they find errors only in those cases when Alice does (or does not, 
respectively) perform the Hadamard operation. Thus, a better strategy for Eve is to randomly apply 
the two Hadamard operations on qubits $P$ and $Q$. She will introduce the same error rate but the 
errors will be equally distributed.
\par
\begin{figure}[pb]
\centerline{\psfig{file=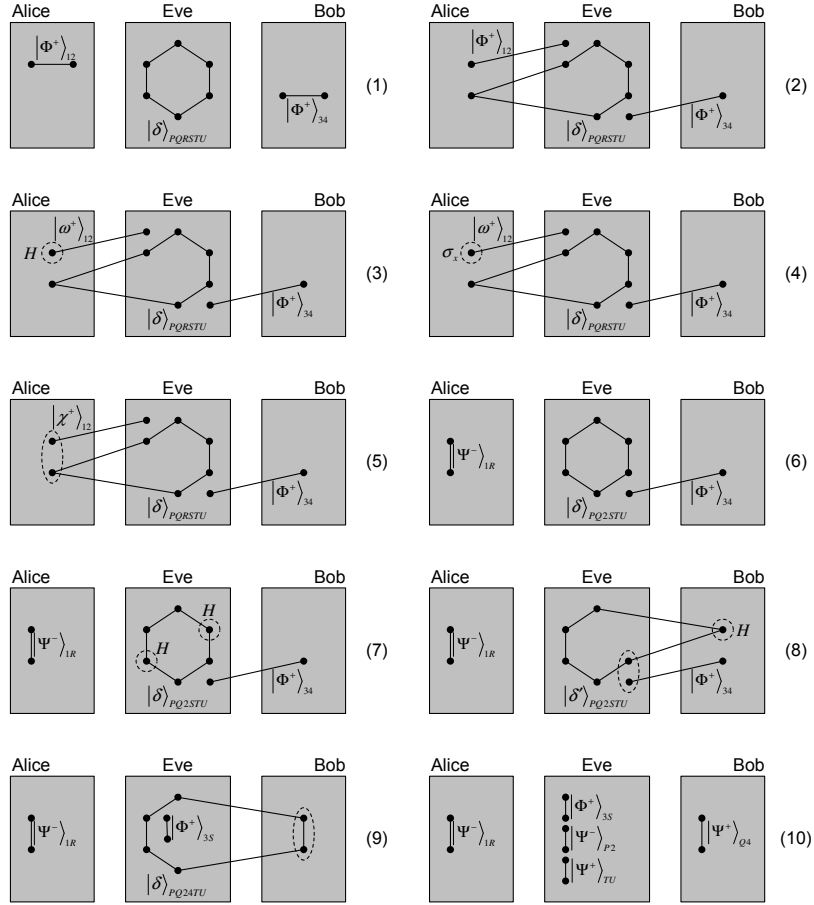,width=0.85\textwidth}}
\caption{An illustration of the attack strategy with delayed measurement on the modified protocol as 
described in section \ref{sec_SecAnal}. Eve waits with her measurement on qubits 2 and $P$ 
until Alice announces whether she applied the Hadamard operation or not.}
\label{pic_ModProtAttack2}  
\end{figure}

Another option for Eve to invert Alice's Hadamard operation is to delay her measurement on the qubit
pairs 2, $P$ and 3, $S$ until Alice publicly announces whether she applied the Hadamard operation or 
not. That means, Eve intercepts qubits 2 and 3 in transit between Alice and Bob and immediately 
forwards qubit $R$ to Alice (cf. picture (2) in figure \ref{pic_ModProtAttack2}). Alice, convinced 
to have received Bob's qubit, applies the Hadamard operation and performs the Bell state measurement 
on qubits 1 and $R$ (cf. pictures (3)-(5) in figure \ref{pic_ModProtAttack2}). Due to entanglement 
swapping qubit 2 is now entangled with qubits $P$, $Q$, $S$, $T$ and $U$, as 
described in eq. \ref{eq_SecAnal_AliceBSM}.
\begin{eqnarray}
\lefteqn{\sigma_A^{(1)} H^{(1)} \ket{\Phi^+}_{12} \otimes \ket{\delta}_{PQRSTU} =} \hspace{30pt} \nonumber \\ 
&\dfrac{1}{2} \Biggl(&
\ket{\Phi^+}_{1R} \otimes \sigma_A^{(P)} H^{(2)} \frac{1}{2\sqrt{2}} \Bigl(
\ket{000000}_{PQ2STU} + \ket{001101}_{PQ2STU} \nonumber \\
& &\quad{}+ \ket{010111}_{PQ2STU} + \ket{011010}_{PQ2STU} + \ket{100110}_{PQ2STU} \nonumber \\
& &\quad{}+ \ket{101011}_{PQ2STU} + \ket{110001}_{PQ2STU} + \ket{111100}_{PQ2STU} \Bigr) \nonumber \\
&+&\ket{\Phi^-}_{1R} \otimes \sigma_A^{(P)} H^{(2)} \frac{1}{2\sqrt{2}} \Bigl(
\ket{000000}_{PQ2STU} - \ket{001101}_{PQ2STU} \nonumber \\
& &\quad{}+ \ket{010111}_{PQ2STU} - \ket{011010}_{PQ2STU} + \ket{100110}_{PQ2STU} \nonumber \\
& &\quad{}- \ket{101011}_{PQ2STU} + \ket{110001}_{PQ2STU} - \ket{111100}_{PQ2STU} \Bigr) \nonumber \\
&+&\ket{\Psi^+}_{1R} \otimes \sigma_A^{(P)} H^{(2)} \frac{1}{2\sqrt{2}} \Bigl(
\ket{000101}_{PQ2STU} + \ket{001000}_{PQ2STU} \nonumber \\
& &\quad{}+ \ket{010010}_{PQ2STU} + \ket{011111}_{PQ2STU} + \ket{100011}_{PQ2STU} \nonumber \\
& &\quad{}+ \ket{101110}_{PQ2STU} + \ket{110100}_{PQ2STU} + \ket{111001}_{PQ2STU} \Bigr) \nonumber \\
&+&\ket{\Psi^-}_{1R} \otimes \sigma_A^{(P)} H^{(2)} \frac{1}{2\sqrt{2}} \Bigl(
\ket{000101}_{PQ2STU} - \ket{001000}_{PQ2STU} \nonumber \\
& &\quad{}+ \ket{010010}_{PQ2STU} - \ket{011111}_{PQ2STU} + \ket{100011}_{PQ2STU} \nonumber \\
& &\quad{}- \ket{101110}_{PQ2STU} + \ket{110100}_{PQ2STU} - \ket{111001}_{PQ2STU} \Bigr)
\Biggr)
\label{eq_SecAnal_AliceBSM}
\end{eqnarray}

Here, Eve does not know in which state her qubits 2, $P$, $Q$, $S$, $T$ and $U$ are because she does
not know Alice's result. But for our further considerations we assume that Alice's result is 
$\ket{\Psi^-}_{1R}$, i.e. Eve is in possession of some state $\sigma_A^{(P)} 
\ket{\delta^\prime}_{PQ2STU}$ similar to her initial state $\ket{\delta}_{PQ2STU}$. Next, Alice 
publicly announces that she applied the Hadamard operation. Having that knowledge Eve applies a $H$ 
operation on qubits 2 and $Q$ in her possession inverting the effect of Alice $H$ operation and 
preparing for Bob's $H$ operation (cf. picture (7) in figure \ref{pic_ModProtAttack2}). Then she 
sends qubit $Q$ to Bob and performs a Bell state measurement on qubits 3 and $S$ which changes the 
state to $\sigma_A^{(P)} \ket{\delta^\prime}_{PQ24TU}$. Finally, Bob inverts the Hadamard operation 
due to Alice's public announcement (cf. picture (8) in figure \ref{pic_ModProtAttack2}). Similar to 
eq. \ref{eq_CESP_ESPO_AS_ES_ESAlice} Bob's Bell state measurement on qubits $Q$ and 4 can be described 
as
\begin{eqnarray}
\sigma_A^{(P)}\ket{\delta^\prime}_{PQ24TU} = \dfrac{1}{2} \Bigl(
&-&\ket{\Phi^+}_{Q4} \otimes \ket{\Phi^-}_{P2} \otimes \ket{\Phi^+}_{TU} \nonumber \\ 
&-&\ket{\Phi^-}_{Q4} \otimes \ket{\Phi^+}_{P2} \otimes \ket{\Phi^-}_{TU} \nonumber \\
&+&\ket{\Psi^+}_{Q4} \otimes \ket{\Psi^-}_{P2} \otimes \ket{\Psi^+}_{TU} \nonumber \\ 
&+&\ket{\Psi^-}_{Q4} \otimes \ket{\Psi^+}_{P2} \otimes \ket{\Psi^-}_{TU} 
\Bigr)
\label{eq_SecAnal_BobBSM}
\end{eqnarray}
It is easy to see that Bob's obtains every Bell state with equal probability of 25\%, i.e. his and 
Alice's results are completely uncorrelated. In case Alice does not apply the Hadamard operation Eve
will introduce the same error rate. Therefore, the probability Eve will be detected is 
\begin{equation}
p = 1 - \biggl( \frac{1}{2} \cdot \frac{1}{4} + \frac{1}{2} \cdot \frac{1}{4} \biggr)^n 
  = 1 - \biggl(\frac{1}{4}\biggr)^n
\end{equation}
which converges much faster to 1 compared to the first strategy described above (cf. eq. 
\ref{eq_ModProt_Error}). 

\section{Conclusion} \label{sec_Concl}

We showed that the protocol presented by Li et al. \cite{LWWSZ06} is open to an attack strategy 
where Eve entangles herself with both legitimate communication parties. Using a special 6-qubit 
state which preserves the correlated results of entanglement swapping an adversary is able to obtain
full information about the key Alice and Bob share in the end. We also discussed an improvement 
where Alice uses a Hadamard transformation on one qubit to secure the protocol. We showed that due 
to the change of the initial state caused by the Hadamard operation an adversary will be detected 
with a high probability.

\section*{Acknowledgements}

We would like to thank Christian Kollmitzer and Oliver Maurhart for fruitful discussions and 
comments. Further we want to thank our reviewer for showing us an interesting new aspect of this 
topic. This work was supported by the Integrated Project SECOQC (Contract No. IST-2003-506813) in 
the Sixth Framework Program of the European Union.

\section*{Appendix. Generation of Eve's Initial State} \label{sec_appendix}

Eve's initial state $\ket{\delta}_{PQRSTU}$ from eq. (\ref{eq_EveInitState}) is rather complex and 
it might not be producible due to today's physical limitations. Nevertheless, we can show that it 
is, in principle, possible for Eve to generate this state using a sequence of Hadamard operations 
and controlled Pauli operations.
\par
First we assume that Eve is in possession of 3 sources emitting Bell states. The state of these 6 
qubits is
\begin{equation}
\ket{\Phi^-}_{PR} \otimes \ket{\Phi^+}_{QS} \otimes \ket{\Phi^+}_{TU}
\end{equation}
Eve performs a Hadamard operation $H$ on qubits $P$ and $Q$ (indicated by the superscripts $(P)$ and 
$(Q)$), which alters the initial state first to
\begin{equation}
H^{(P)} \ket{\Phi^-}_{PR} \otimes \ket{\Phi^+}_{QS} \otimes \ket{\Phi^+}_{TU} 
\end{equation}
and further to
\begin{eqnarray}
\lefteqn{H^{(P)} \ket{\Phi^-}_{PR} \otimes H^{(Q)} \ket{\Phi^+}_{QS} \otimes \ket{\Phi^+}_{TU}=} \hspace{30pt} \nonumber \\ 
&\frac{1}{2} \Bigl(&
  \ket{\Phi^+}_{PR} \otimes \ket{\Phi^-}_{QS} \otimes \ket{\Phi^+}_{TU} 
+ \ket{\Phi^+}_{PR} \otimes \ket{\Psi^+}_{QS} \otimes \ket{\Phi^+}_{TU} \nonumber \\
&-&\ket{\Psi^-}_{PR} \otimes \ket{\Phi^-}_{QS} \otimes \ket{\Phi^+}_{TU}
-\ket{\Psi^-}_{PR} \otimes \ket{\Psi^+}_{QS} \otimes \ket{\Phi^+}_{TU}
\Bigr)
\end{eqnarray}
Then Eve applies a $\sigma_z$ operation on qubit $Q$ if qubits $P$ and $R$ are in the Bell state 
$\ket{\Phi^+}_{PR}$. Otherwise the identity operator is applied. This action can be described as
\begin{eqnarray}
\oP{\Phi^+}{\Phi^+} \otimes \sigma_z \otimes I \otimes I \otimes I
&+&\oP{\Phi^-}{\Phi^-} \otimes I \otimes I \otimes I \otimes I \nonumber \\
{}+\oP{\Psi^+}{\Psi^+} \otimes I \otimes I \otimes I \otimes I
&-&\oP{\Psi^-}{\Psi^-} \otimes I \otimes I \otimes I \otimes I 
\end{eqnarray}
and it changes the state to
\begin{eqnarray}
&\dfrac{1}{2} \Bigl(&
  \ket{\Phi^+}_{PR} \otimes \ket{\Phi^+}_{QS} \otimes \ket{\Phi^+}_{TU}
+ \ket{\Phi^+}_{PR} \otimes \ket{\Psi^-}_{QS} \otimes \ket{\Phi^+}_{TU} \nonumber \\
&+&\ket{\Psi^-}_{PR} \otimes \ket{\Phi^-}_{QS} \otimes \ket{\Phi^+}_{TU}
+ \ket{\Psi^-}_{PR} \otimes \ket{\Psi^+}_{QS} \otimes \ket{\Phi^+}_{TU}
\Bigr)
\end{eqnarray}
Next Eve performs a $\sigma_x$ operation on qubit $P$ and a $\sigma_z$ on qubit $T$ if qubits $Q$ 
and $S$ are in the state $\ket{\Phi^-}_{QS}$. Alternatively, she performs a $\sigma_z$ operation on 
qubit $P$ and a $\sigma_x$ operation on qubit $T$ if qubits qubits $Q$ and $S$ are in the state 
$\ket{\Psi^+}_{QS}$. If qubits $Q$ and $S$ are in the state $\ket{\Psi^-}_{QS}$ Eve applies a 
$i\sigma_y$ operation on both qubits $P$ and $T$. These three controlled operation can be stated as
\begin{eqnarray}
           I \otimes I \otimes \oP{\Phi^+}{\Phi^+} \otimes I \otimes I
&+&\sigma_x  \otimes I \otimes \oP{\Phi^-}{\Phi^-} \otimes \sigma_z \otimes I \nonumber \\
{}+\sigma_z  \otimes I \otimes \oP{\Psi^+}{\Psi^+} \otimes \sigma_x \otimes I 
&+&i\sigma_y \otimes I \otimes \oP{\Psi^-}{\Psi^-} \otimes i\sigma_y \otimes I 
\end{eqnarray}
Using this operator Eve is able to bring the 6 qubits into the desired state from eq. 
(\ref{eq_EveInitState}).

\end{document}